\documentclass[printer]{aa}
  
\usepackage{graphicx}
\usepackage{natbib}
\newcommand{\be}{\begin{eqnarray}}
\newcommand{\ee}{\end{eqnarray}}

\newcommand {\nbodypp}{\textsc{\mbox{nbody6\raise.4ex\hbox{\tiny++}}}}

\newcommand {\Msun} {\mbox{M$_{\odot}$}}

\begin{document}

\title{Universality of young cluster sequences}
\author{S. Pfalzner} 
\institute{I. Physikalisches Institut, University of Cologne, Z\"ulpicher Str. 77, 50937 Cologne, Germany}
\date{}

\abstract
{}
{Most stars do not form in isolation but as part of a cluster comprising anywhere between a few dozen to several
million stars with stellar densities ranging from 0.01 to several 10$^5$ \Msun pc$^{-3}$. The majority of these
clusters dissolve within 20 Myr.  The general assumption is that clusters are born more or less over this entire 
density range.} 
{A new analysis of cluster observations is presented.} 
{It demonstrates that, in fact, clustered star 
formation works under surprisingly tight constraints with respect to cluster size and density.} 
{The observed multitude of cluster densities simply results from snapshots of two sequences evolving in time along 
pre-defined tracks in the density-radius plane. This implies that the cluster size can actually be used to determine 
its age.}

\maketitle

\section{Introduction}

The very early phase of star formation is still poorly understood because most stars form in dense clusters embedded 
within giant molecular clouds \cite{lada:03,pudritz:02} often only observable at infrared wavelengths. If the clusters 
consist of a thousand stars or more, they also contain massive O stars that drive the expulsion of gas from the 
cluster via stellar winds, ionisation and supernovae of early-type stars. This gas expulsion brings the clusters out 
of dynamical equilibrium, eventually exposing them and leading to a cluster expansion where the majority of 
stars become unbound \cite{hills:80,goodwin:06,baumgardt:07,bastian:08}, eventually turning into field stars.  
Observations support this picture because young ($\sim$ 1 Myr) clusters are usually smaller than older ($\sim$ 20 Myr) 
clusters, but a systematic and quantitative picture of the cluster expansion itself has so far been lacking. It is 
this evolution of exposed clusters over the first 20 Myr that we address here. It is a time-span of special interest 
because it is precisely the period when 
planetary systems are developing.

The observed densities of young star clusters range from less than 0.01 to several 10$^5$ 
\Msun pc$^{-3}$ as Fig.~1 shows (a list of cluster properties is given in Table 1 in the online material: 
Figer 2008, Wolff et al. 2007, Borissova et al. 2008), where the determination methods of the masses and radii from the literature 
sources are also explained). Here only clusters more massive than 1000 \Msun (massive clusters) are included, since 
the size and density of clusters containing only a few dozen stars are poorly determined.  The wide variety of 
cluster densities has lead to the general assumption that clusters are also formed over this entire density range. As 
a result of   the gas expulsion, a rapid dissolution of the cluster because of their super-virial velocity dispersion  
\cite{hills:80,kroupa:05,goodwin:06,gieles:08} is then expected.

\section{Results}

\begin{figure}[t]
\resizebox{\hsize}{!}{\includegraphics[angle=-90]{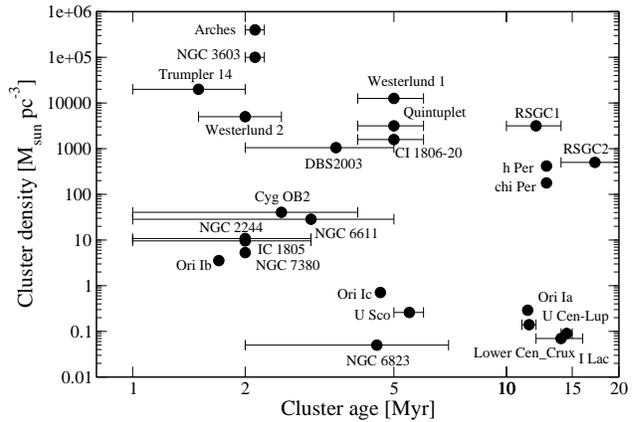}}
\caption{Cluster density as a function of age for clusters more massive than 10$^3$ \Msun.
The values were taken from Figer (2008), Wolff (2007), and Borissova et al. (2008) and
references therein. See Table 1 in online material as well for a discussion of the error bars.}
\label{fig:cluster_age}
\end{figure}

When looking closely at the same clusters as in Fig.~1 in a different way by plotting instead the density as 
a function of the cluster 
radius $R$ (definitions for the individual data points are given in the online material) leads to the remarkably 
simple structure in Fig.2. Apart from clusters existing in two groups, what this plot unambiguously 
shows is that these evolve along two well-defined tracks in the density-radius plane, strongly suggesting a  
bi-modal cluster evolution. Moreover, as can be seen in Fig.~2, the two classes of clusters - 
in the following called starburst and leaky clusters (the terminology will be explained later) – 
each start from unique points in the mass-radius plane and develop at approximately the 
same speed along these two tracks. All clusters shown originate from one of these birth points in mass and radius.

This is not so surprising for starburst clusters since their relatively similar mass has already led to 
speculations that they might have a common origin. For leaky clusters, however, the prevailing view is that 
young clusters form more or less at arbitrary density and somehow dissolve afterwards on timescales of ~20 Myrs. 
However, the new representation of already known cluster parameters in Fig. 2 shows  the common 
history of these clusters for the first time. The single track and sequential nature of the data points means that, 
contrary to the usual viewpoint, these clusters also start out with the same mass density and radial extent 
(and therefore mass) and all expand in the same way. 

Here a word about determinating of radii and densities in the different sources is nessesary. 
All starburst cluster data, apart from that for DBS 2003, 
were taken from Figer (2008). The data for DBS 2003 were taken from Borissova et al. (2008).  
In Figer (2008) the radius gives the average projected separation from the centroid position. 
The total cluster mass is given by extrapolating down to a lower mass cut-off of 1 solar mass, assuming a Salpeter 
initial mass function and an upper mass cut-off of 120 solar masses (exceptions from this method are given 
in Figer 2008). The total mass of the cluster was determined by assuming that the stars in the mass 
range 6{\bf -}12 \Msun\ constitute 5.5\% of the total mass of the complete population of stars spanning the 
range 0.1{\bf -}100 \Msun. They compared their values with other mass-determining methods by Slesnick 
et al. (2002) and Weidner \& Kroupa {\bf (2006)} and found good agreement.  In Borissova et al. the radius was 
defined by distance where the density profile exceeds twice the standard deviation of the surface density in the 
surrounding field. All values for the leaky clusters were taken from Wolff {\bf et al.} (2007). There the radius 
was determined as the median distance in degrees of the B stars in the sample from the centre 
of the cluster, so that, although not all cluster sizes were determined exactly in the same way, for each subset - 
the starburst clusters and the leaky clusters - the cluster radii were determined fairly consistently. 
The density was in all cases defined as $\rho = 3 M_c/4 \pi R^3$, so it is extremely unlikely that the result
is an artefact of the applied method.

\begin{figure}[t]
\resizebox{\hsize}{!}{\includegraphics[angle=-90]{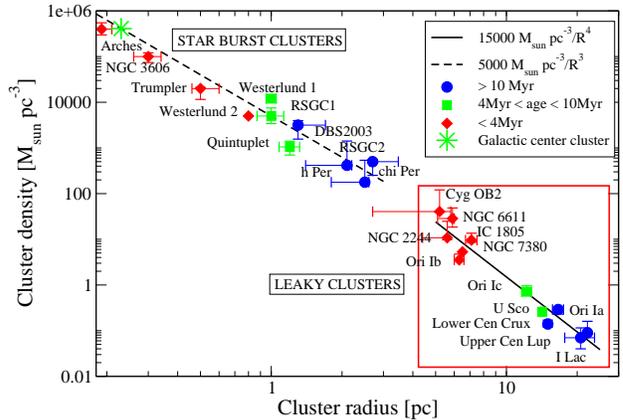}}
\caption{{Cluster density as a function of cluster size for clusters more massive than 10$^3$ 
\Msun.
The values were taken from Figer (2008), Wolff et al. (2007) and Borissova et al. (2008) and
references therein. See although Table 1 in online material.}}
\label{fig:cluster_rad}
\end{figure}

Quantitative analysis of Fig.2 shows that, as the leaky clusters expand, they do not simply diffuse (which would give 
an $R^{-3}$-dependence), but the cluster density decreases as $R^{-4}$. This means that the cluster loses  mass via 
various processes such as outgassing, stellar mass loss, tidal evaporation, and escapers created through three-body 
encounters close to its centre. The relative importance of these mass-loss processes will need clarification in the 
future. As a direct consequence of the common history of these clusters, one can now determine the cluster size as a 
function of cluster age and finds a $t_c^{0.6-0.7}$-dependence. As demonstrated in Fig.~3, 
the expansion velocity is $\sim$ 2pc/Myr in the early phases and drops to $\sim$1pc/Myr at cluster ages 
$\sim$20 Myr. In Fig.3 it is shown that comparing simulations of clusters expanding due to gas expulsion showing similar 
expansion velocity 
\cite{olczak} indicates that $\sim$ 75\% of the gas is expelled, though more detailed studies are required to 
confirm this. However, this would already suggest that all leaky clusters shown in Fig.2 formed with the 
approximately the same star formation efficiency of $\sim$ 25\%.

It is thought that clusters and associations are formed with mass functions that are powerlaws with index -2.  
However, the results described here imply that the leaky clusters form in a narrow mass range (3.6 $<$ log $M_c <$ 4.4) and develop
quickly within 20 Myrs to lower masses (3.3 $<$ log $M_c <$ 3.6). The small sample size currently does not allow
further constriction of the primordial mass function for loss clusters. 

The situation is simpler for the starburst clusters. They all appear to be born with a cluster density of at least 
10$^5$ \Msun pc$^{-3}$  (red data points in the top left corner of Fig.~2) and then simply diffuse without further mass 
loss, i.e. a $R^{-3}$–dependence.   In fact the expansion proceeds more or less linearly in time, i.e. as $R$ 
$\sim t_c$ (see Fig.~3). Having established a common history, one can easily deduce an expansion velocity of 0.1-0.2 
pc/Myr. From this expansion velocity, it should be possible to determine the star formation efficiency through 
numerical simulations in the future. A possible reason for the expansion velocity in star-burst clusters to be lower 
than in 
leaky clusters is that the potential well is too deep for the ionised  gas to leave the cluster quickly 
\cite{kroupa:08}. This means a much higher star formation efficiency for starburst clusters than for the massive 
clusters. 

Fig.~2 shows an additional feature, namely the location of the Galactic center clustre. Since the Galactic centre 
harbours a massive black hole, one would expect its dynamics to differ considerably from starburst clusters. 
However, it lies directly on the starburst cluster track close to the data point of Arches. Given that the Galactic 
centre cluster is older than the Arches cluster, it follows that it must have started out at a higher initial mass 
and/or expanded more slowly.

\begin{figure}[t]
\resizebox{\hsize}{!}{\includegraphics[angle=-90]{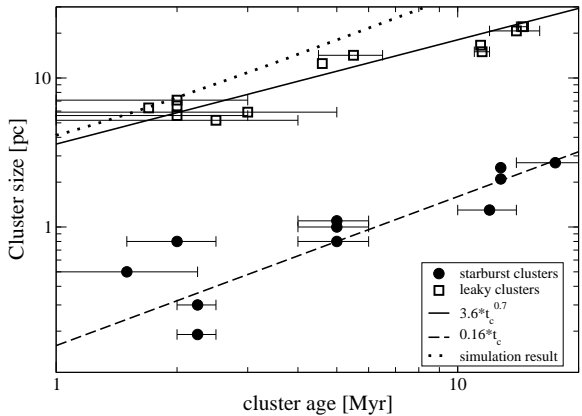}}
\caption{Cluster size as a function of cluster age for the same clusters in 
Fig.~\ref{fig:cluster_rad}.
The squares correspond to the leaky clusters and the circles to the starburst clusters.
The dotted line indicates simulation results by Olczak (2009) for an ONC-like cluster with 75\% gas expulsion.
The drawn and the dashed lines are not fits but just there to guide the eye.}
\label{fig:cluster_size_age}
\end{figure}

\begin{figure}[t]
\resizebox{\hsize}{!}{\includegraphics[angle=-90]{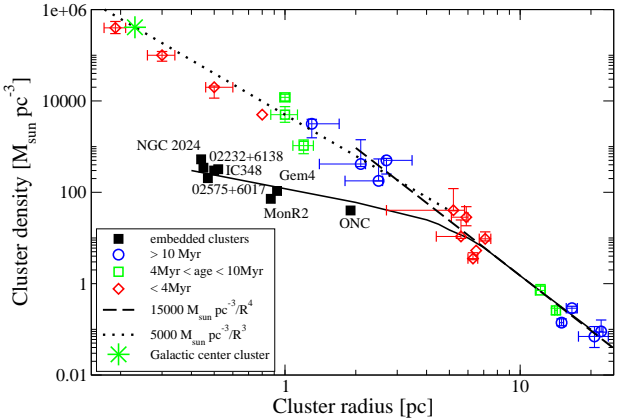}}
\caption{Same as Fig.~\ref{fig:cluster_rad} but here the cluster density of embedded clusters with more than observed 
200 members are also shown. The values were taken from Lada \& Lada 2003.}
\label{fig:cluster_rad_emb}
\end{figure}

If the above systematics is true, then no clusters should exist with $R >$ 2 pc ages ages $<$ 4Myr that possess masses 
higher than several 10$^4$ \Msun. An interesting consequence of the sequential nature of cluster development is that 
the radius of the cluster can 
actually be used to determine its age. Cluster age determination is notoriously problematic, especially for very 
young clusters, where the uncertainty of the tracks of pre-main sequence star development can lead to very different 
age estimates. The density-radius relation defined here might be an easier and potentially more accurate method for 
cluster age determination. Clusters with a radius between 2-4 pc pose a special problem. Only the distinction between 
very young leaky clusters ($<$ 4Myr) and old starburst clusters ($>$ 10 Myr) will require additional information due 
to their similar radii. However, as one
can expect that in older starburst clusters some red supergiants are present, telling them apart is probably not too 
difficult.

How do embedded clusters fit into this picture? Fig.~4 again shows the cluster density as a function of size but also 
includes embedded clusters. The embedded clusters were selected from the data by Lada \& Lada (2003) with the 
restriction that at least 200 cluster stars were detected.  The embedded clusters are not located on the main 
evolutionary sequence but form their own side-arm off the leaky cluster sequence below its extension to younger 
ages. It joins up when the embedded clusters become exposed after having reached a size of $\sim$ 2-4 pc  at an age 
of  $>$ 1 Myr. A further argument for the sequential nature of the cluster development is that the ONC is likely to 
be expanding \cite{kroupa:01},  hence evolving along a time-density track. In other words, a relatively sparse cluster 
like Upper Cen Lup must once have been in a state similar to the Orion Nebular Cluster now – the densest nearby 
cluster  (see Fig.~{\bf 4}). Note as well that Ori Ib, 
Ori Ic, and Ori Ia form a temporal sequence of once leaky clusters with initial densities as high as the ONC. 
The initial stellar mass content of leaky clusters at the end of the embedded phase is at least 3500 \Msun . 
with a radius exceeding $\sim$ 4pc, 

One would expect a similar side-arm for the starburst clusters, but so far no such embedded starburst clusters have 
been observed. However, this is not surprising since they would be very small ($R <$ 0.1 pc) and located either close 
to the Galactic centre (at a distance of $\sim$ 8kpc) or inside spiral arms within the Galactic plane. In addition,
starburst clusters are intrinsically rare, and the embedded phase very short, so not many are expected to be found.

\section{Discussion and conclusions}

The above results imply that star formation occurs only under an extremely limited set of conditions, and may 
therefore require a fundamental revision of its theories in our Galaxy, possibly including our own solar system. 
The ``leaky clusters'' could also be known as ``mass-loss clusters",
as this is what distinguishes them from starburst clusters.''. For 
this leaky cluster density regime, recent simulations \cite{kuepper:08} have found that the dynamics of embedded 
clusters show a strong correlation between the kinetic energy of the cluster stars and that of stars leaving the 
cluster, leading to a common sequence. This prediction might well be linked to the results described herein.   

The embedded cluster side-arm implies that many of the older clusters (like Ori Ia-c, Upper Sco, I Lac, etc.) must 
have gone through a phase like the ONC is in today. Thus the ONC can serve as model cluster for both the earlier 
phase of older leaky clusters and the future of the younger embedded clusters. Results for the ONC 
therefore serve as a template for a multitude of other clusters. For example, observations and theory indicate that 
the central density in the ONC is so high that encounters between young disc-surrounded stars might affect the 
protoplanetary discs \cite{herbig:86,pfalzner:aa06} and changes in the disc properties are likely to have direct 
consequences on the ongoing planet formation process. 

Finally the proposed reordering immediately raises the question of the origin of these two distinct cluster 
groupings. The critical parameter is possibly the higher density, but could equally be the density-temperature ratio 
or different types of turbulence, etc. The obvious task for the near future is to understand why these specific 
conditions are favoured in our Galaxy and whether the same applies in other spiral galaxies, too.

\scriptsize
\bibliographystyle{apj}

\begin{thebibliography}{}

\small


\bibitem[Baumgardt \& Kroupa \- 2007]{baumgardt:07}
Baumgardt \& Kroupa 2007, \mnras, 380, 1589.

\bibitem[Bastian et al.  \- 2008]{bastian:08}
Bastian et al. 2008, \mnras, 389, 223.

\bibitem[Borissova et al. \- 2008]{borissova:08}
Borissova, J., Ivanov, V.D., Hanson, M.M., Georgiev, L., Minniti, D., Kurtev, R.,
Geisler, D. 2008, \aap, 488, 151.

\bibitem[de Gris et al. 2003]{degris:03}
de Grijs, R., Lee, J.T., Mora Herrera, M. C., Fritze-v. Alvensleben, U., Anders, P.,
2003, New Astronomy, 8, 155.

\bibitem[Figer 2008]{figer:08}
Figer, D.F., 
2008, "Massive Stars as Cosmic Engines", Proc. of  Intern. Astron. Union, IAU Symposium, 250, 247.

\bibitem[Gieles \& Bastian 2008]{gieles:08}
Gieles, M., Bastian, N., 
2008, A\&A 482, 165.

\bibitem[Goodwin \& Bastian 2006]{goodwin:06} 
Goodwin, S. P., Bastian, N., 
2006, MNRAS 373, 752.

\bibitem[Herbig \& Terndrup 1986]{herbig:86}
Herbig, G. H., Terndrup, D. M., 
1986, ApJ, 307, 609.

\bibitem[Hills 1980]{hills:80}
Hills, J.G., 
1980, ApJ 235, 987

\bibitem[Kroupa 2005]{kroupa:05} 
Kroupa, P., 
2005, Proc. of "The Three-Dimensional Universe with Gaia" (ESA SP-576)., Paris-Meudon, ed.: C. Turon, K.S. 
O'Flaherty, M.A.C. Perryman, p.629

\bibitem[Kroupa et al. 2001]{kroupa:01} 
Kroupa, P., Aarseth, S., Hurley, 2001, \mnras, 321, 699. 

\bibitem[Kroupa 2008]{kroupa:08}
Kroupa, P., 
2008, in Proc. of ``Mass Loss from Stars and the 
Evoltution of Stellar Clusters'', ASP Conf. Ser. 388, p. 271, eds. A. de Koter, L. Smith, R. Waters, 
 Astron. Soc. Pac.,  San Francisco.

\bibitem[K\"upper et al. 2008]{kuepper:08}
K\"upper, A.H., Kroupa, P., Baumgardt, H., 
2008, MNRAS 389, 889.

\bibitem[Lada \& Lada 2003]{lada:03}
Lada, C.J., Lada, E.A., 
Annu.Rev. Astron. Astrophysics, 2003, 41, 57.

\bibitem[Olczak 2008]{olczak}
Olczak, C., private communication

\bibitem[Pfalzner et al. 2006]{pfalzner:aa06}
Pfalzner, S., Olczak, C., Eckart, A.,  
2006,  A\&A, 454, 811.

\bibitem[Pudritz 2002]{pudritz:02}
Pudritz, R.E., 2002, Science, 295, 68

\bibitem[slesnick et al. 2002]{slesnick:2002}
Slesnick et al. 2002, ApJ, 576, 880

\bibitem[Wolff et al. 2007]{wolff:07}
Wolff, S. C., Strom, S. E., Dror, D., Lanz, L., Venn, K., 
2007, Astron. J., 132, 749.

\bibitem[Weidner \& Kroupa 2006]{weidner:2006}
Weidner \& Kroupa, 2006, MNRAS, 365, 1333



\end{thebibliography}

\onecolumn

\large
{\bf Online material}

\normalsize

\begin{table}
\begin{center}
\begin{tabular}{l|*{5}{c}}
Identification & distance & age   &  log($M_c$)  & size  & log($\rho_c$)      \\[0.5ex]
               & [pc]     & [Myr] &  [\Msun]    & [pc]  & [\Msun pc$^{-3}$] \\[0.5ex]
\hline
\\[-2ex]
Arches$^1$         &  8  $^{+1}_{-1}$       & 2-2.5  & 4.3   & 0.19 $^{+0.03}_{-0.03}$&  5.6   $^{+1}_{-1}$\\[0.5ex]
NGC 3603$^1$       &  7.6$^{+1}_{-1}$       & 2-2.5  & 4.1   & 0.3  $^{+0.04}_{-0.04}$&  5.0   
$^{+0.1}_{-0.1}$\\[0.5ex]
Trumpler 14$^1$    &  2.8$^{+0.6}_{-0.2}$   & $$<$ 2$& 4.0   & 0.5  $^{+0.1}_{-0.04}$ &  4.3   
$^{+0.05}_{-0.3}$\\[0.5ex]
Westerlund 1$^1$   &  3.55$^{+0.17}_{-0.01}$& 4-6    & 4.7   & 1.0  $^{+0.05}$        &  4.1   
$^{+0.06}_{-0.03}$\\[0.5ex]
Westerlund 2$^1$   &  2.8                   &1.5-2.5 & 4.0   & 0.8                    &  3.7   \\[0.5ex]
RSGC 1$^1$         &  5.83$^{+1.9}_{-0.78}$ & 10-14  & 4.5   & 1.3  $^{+0.04}_{-0.04}$&  3.5   
$^{+0.1}_{-0.3}$\\[0.5ex]
Quintuplet$^1$     &  8$^{+1}_{-1}$         & 4-6    & 4.3   & 1.0  $^{+0.13}$        &  3.7   
$^{+0.7}_{-0.2}$\\[0.5ex]
DBS2003$^2$        &  7.9$^{+1.2}_{-1}$     & 2-5    & 3.8   & 1.2  $^{+0.18}_{-0.5}$ &  3.05  
$^{+0.5}_{-0.3}$\\[0.5ex]
RSGC 2$^1$         &  5.83$^{+1.91}_{-0.78}$& 14-21  & 4.6   & 2.7  $^{+0.77}_{-0.77}$&  2.7   
$^{+0.05}_{-0.03}$\\[0.5ex]
$\chi$ Per$^3$     &  2.34$^{+0.1}_{-0.5}$  & 12.8   & 4.1   &  2.5 $^{+0.1}_{-0.7}$  &  2.61  
$^{+0.5}_{-0.05}$\\[0.5ex]
h Per$^3$          &  2.34$^{+0.1}_{-0.5}$  & 12.8   & 4.2   &  2.1 $^{+0.1}_{-0.7}$  &  2.25  
$^{+0.5}_{-0.05}$\\[0.5ex]
CYg OB2$^3$        &  1.74$^{+0.2}_{-0.5}$  & 1-4    & 4.4   &  5.2 $^{+0.06}_{-2.5}$ &  1.61  
$^{+0.02}_{0.4}$\\[0.5ex]
NGC 6611$^3$       &  1.995$^{+0.01}_{-0.25}$& 1-5   & 4.4   &  5.9 $^{+0.1}_{-0.8}$  &  1.45  
$^{+0.22}_{0.11}$\\[0.5ex]
NGC 2244$^3$       &  1.88$_{-0.4}$         & 1-3    & 3.9   & 5.6  $_{-1.2}$         &  1.03  $^{+0.33}$\\[0.5ex]
IC 1805$^3$        &  2.34$^{+0.1}_{-0.1}$  & 1-3    & 4.2   &  7.1 $^{+0.3}_{-0.3}$  &  0.98  $^{+0.03}$\\[0.5ex]
Ori Ib$^3$         &  0.363$^{+0.2}_{-0.2}$ & 1.7    & 3.6   &  6.3 $^{+0.3}_{-0.3}$  &  0.55  
$^{+0.11}_{-0.02}$\\[0.5ex]
NGC 7380$^3$       &  3.73                  & 2      & 3.8   &  6.5                   &  0.72  \\[0.5ex]
Ori Ic$^3$         &  0.398$^{+0.2}_{-0.2}$ & 4.6    & 3.8   & 12.5 $^{+0.6}_{-0.6}$  &  -0.15 
$^{+0.12}_{-0.01}$\\[0.5ex]
Ori Ia$^3$         &  0.380$^{+0.2}_{-0.2}$ & 11.4   & 3.7   & 16.6 $^{+0.9}_{-0.9}$  &  -0.54 
$^{+0.05}_{-0.09}$\\[0.5ex]
U Sco$^3$          &  0.144$^{+0.003}_{-0.003}$& 5-6 & 3.5   & 14.2 $^{+0.2}_{-0.2}$  &  -0.59 
$^{+0.06}_{-0.01}$\\[0.5ex]
Lower Cen-Crux$^3$ &  0.116$^{+0.002}_{-0.002}$& 11-12 & 3.3 & 15.0 $^{+0.3}_{-0.3}$  &  -0.85 
$^{+0.05}_{-0.01}$\\[0.5ex]
Upper Cen-Lup2$^3$ &  0.142$^{+0.002}_{-0.002}$& 14-15 & 3.6 & 22.1 $^{+0.4}_{-0.4}$  &  -1.05 
$^{+0.01}_{-0.01}$\\[0.5ex]
I Lac 2$^3$        &  0.368$^{+0.06}$          & 12-16 & 3.4 & 20.7 $^{+3}_{-3}$      &  -1.15 
$^{+0.2}_{-0.17}$\\[0.5ex]
\end{tabular}
\caption{Properties of clusters and associations more massive than 10$^3$ \Msun\ in the Galaxy
{$^1$ \cite[and references therein]{figer:08}, $^2$ \cite{borissova:08},  
$^3$\cite[and references therein]{wolff:07}. The errors in age result from (a) their deduction from theoretical 
pre-main-sequence star tracks for young 
stars, which show large discrepancies; (b) the age of the cluster is found by averaging over the so deduced ages of 
the single stars; and (c) the age determination depends on the knowledge of the distance to the cluster - a parameter 
that is often poorly constrained. In cases where there is no error bar this does not mean that the age is accurately 
known, but rather that the authors did not give any error value. Usually there are no error values quoted for the mass or the                  
density of 
the cluster, nor is the cluster size consistently defined for the different clusters. An error of a factor of 2 is 
probably realistic and can be expected to be even larger for the more distant clusters, so for this and the following 
figures one should always keep in mind that the errors of these observed values are rather large.}
\label{table:nacc1}}
\end{center}
\end{table}

\begin{table}
\begin{center}
\begin{tabular}{r|*{5}{c}}
Identification & distances&  $M_c$   & size  & log($\rho_c$)  \\[0.5ex]
               & [pc]     & [\Msun] & [pc]  & [\Msun pc$^{-3}$] \\[0.5ex]
\hline
\\[-2ex]
02232 + 6138$^1$   & 2.4  & 130   & 0.91 & 2.23  \\[0.5ex]
02575 + 6017$^1$   & 2.4  & 150   & 1    & 2.17  \\[0.5ex]
IC 348$^1$         & 0.32 & 160   & 1    & 2.20  \\[0.5ex]
ONC $^1$           & 0.45 & 1100  & 3.8  & 1.30  \\[0.5ex]
NGC 2024  $^1$     & 0.4  & 180   & 0.88 & 2.42  \\[0.5ex]
MonR2  $^1$        & 0.8  & 340   & 1.85 & 1.72  \\[0.5ex]  
\end{tabular}
\caption{Properties of embedded clusters containing more than 200 observed stars.\newline 
$^1$ \cite[][and references therein]{lada:03}
\label{table:nacc2}}

\end{center}
\end{table}

\end{document}